\documentclass[11pt]{article}
\usepackage{amssymb}
\usepackage{amsfonts}
\usepackage{amsmath}
\usepackage{geometry}
\usepackage{graphicx}
\usepackage{subfigure}
\usepackage[singlespacing]{setspace}

\begin{document}

\title{Quantum Key Distribution \\
Highly Sensitive to Eavesdropping}
\author{{Stephen Brierley} \\
{\normalsize Department of Mathematics, University of York}\\
{\normalsize Heslington, York, UK YO10 5DD}\\
{\normalsize \texttt{sb572@york.ac.uk}}}
\date{{\normalsize October 2009}}
\maketitle

\begin{abstract}
We introduce a new quantum key distribution protocol that uses $d$-level
quantum systems to encode an alphabet with $c$ letters. It has the property
that the error rate\ introduced by an intercept-and-resend attack tends to
one as the numbers $c$ and $d$ increase. In dimension $d=2$, when the
legitimate parties use a complete set of three mutually unbiased bases, the
protocol achieves a quantum bit error rate of $57.1\%$. This represents a
significant improvement over the $25\%$ quantum bit error rate achieved in
the BB84 protocol or $33\%$ in the six-state protocol.
\end{abstract}

\section{Introduction}

By sharing a random string of numbers, two parties can encrypt a message in
such a way that it appears completely random to an eavesdropper. The
\textquotedblleft one-time pad\textquotedblright\ is an unbreakable method
of encryption provided the string is truly random and only used once. The
problem comes in having sufficiently many strings, called keys, with which
to encrypt all messages you wish to send. This is called the key
distribution problem.

By allowing the security of the key distribution protocol to be \emph{%
computationally} impossible rather than unconditional, several ingenious
methods to distribute keys have been developed. Assuming that an
eavesdropper does not posses an infinitely large computer, cryptographic
systems can make use of mathematical problems that are very hard to solve.
For example, there is no known efficient algorithm to factorize large
integers into a product of primes (used in the Rivest-Shamir-Adleman
algorithm \cite{rivest+78}) or to compute the discrete logarithm (used in
the Diffe-Hellman-Merkle key exchange \cite{diffie+76, hellman+80}).
However, solving these mathematical problems is only difficult, not
impossible, so that the security of such public key protocols relies on the
lack of future developments in mathematics and technology.

In 1970, Wiesner proposed a totally new approach to cryptography \cite%
{weisner70} that was then developed by Bennett and Brassard: they presented
a key distribution protocol \cite{bennet+84}, now known as BB84, that uses
properties of quantum systems to ensure its security. This protocol allows
two parties, Alice and Bob, to distribute a key such that anyone who
attempts to listen in on the quantum signals can, in theory, be detected.
The eavesdropper, Eve, is constrained by the physical laws of quantum
mechanics. She cannot perform a measurement without introducing a
disturbance (Heisenberg's uncertainty principle), copy states (no cloning)
or split the signal, since it consists of single photons or particles. Other
protocols such as Ekert's \cite{ekert91} use entangled particles in such a
way that Eve essentially introduces hidden variables destroying the quantum
correlations. It is possible to prove that these quantum key distribution
(QKD) protocols are secure against all future technological and mathematical
advances\footnote{%
Except possibly a new theory of physics that allows operations beyond
quantum mechanics (c.f. Popescu-Rohrlich boxes \cite{popescu+94}).} \cite%
{mayers01, shor+00}.

When attempting to implement a QKD protocol a key factor in determining its 
\emph{practical} success is the error rate introduced by Eve: if it is
small, her presence may be masked by the system noise. This error rate thus
determines the level of technology required to implement the protocol and
the distance\ over which Alice and Bob can establish a secure key. We will
present a protocol that extends the one proposed by Khan et al. \cite%
{khan+09}. The new\ approach ensures that eavesdropping causes a large error
rate and therefore, from an experimental point of view, offers a
modification that could improve the implementation of existing QKD
technology.

The new protocol allows Alice and Bob a great deal of freedom: the elements
of the key that they form can be taken from an alphabet of arbitrary size,
and encoded using any bases of $\mathbb{C}^{d}$. It is equivalent to the
protocol presented in \cite{khan+09} when Alice and Bob use a two-letter
alphabet and corresponds to the SARG protocol \cite{scarani+04}\ when\ in
addition, they use two-dimensional quantum systems.

In order to better understand the freedom in the choice of bases used by all
three parties, Alice, Bob and Eve, we will introduce a measure of distance
between two bases and show how this relates to the error rate. It gives a
simple interpretation of the optimal setup for all parties: Alice and Bob
should use a set of $c$ bases, $\mathcal{S}$, that are as far apart as
possible; whilst Eve should choose her basis, $\mathcal{E}$, so it minimises
the average distance between $\mathcal{E}$ and the elements of $\mathcal{S}$%
. The conclusion then is that for the legitimate parties, the optimal
settings correspond to so called \emph{mutually unbiased (MU) bases} or
complementary observables. MU bases have the property that a measurement in
any one of the bases reveals no information about the state in all of the
other bases; and have been used before in other QKD protocols \cite%
{bennet+84, bourennane+01, cerf+02}.

The paper is organised as follows. In Sec. \ref{general scheme}, we will
introduce a key distribution protocol that encodes a $c$-letter alphabet
using quantum systems of dimension $d.$\ In Sec \ref{Eve's strategy}, we
will examine the effect of an eavesdropper by calculating two error rates
that allow the legitimate parties to detect Eve's intercept-and-resend
attack. Sec. \ref{distance} will show how one of these error rates can be
understood as a measure of the distance between the bases used by all three
parties. We will consider some examples of specific sets of bases in Sec. %
\ref{specific bases}. In Sec. \ref{Example}, we compare this new protocol to
the six-state protocol in an experimental setting and consider a general
method of implementing the protocol for any choice of $c$ and $d$. Finally,
we summarise the results and compare the new protocol to existing quantum
key distribution methods in Sec. \ref{conclusion}.

\section{General form of the Protocol \label{general scheme}}

In quantum cryptography, there are two legitimate parties who wish to
establish a shared sequence of letters from an alphabet such as a string of
zeros and ones. Typically, these two parties have different roles: Alice
prepares and sends quantum states, and Bob performs measurements on the
states he receives and records the outcomes. At the end of this quantum part
of the protocol, the two parties then exchange information via a classical
communication channel. A third party, Eve, attempts to gain information
about some or all of the shared key without being detected. Eve can perform
any operation allowed by quantum mechanics and can listen in on the
classical part of the communication without being detected. We also assume
that she has access to a high level of technology so that she can hide
behind any system noise by replacing parts of the implementation by better
components. The aim is to find protocols and implementations such that Eve
is easily detected.

We begin by presenting a new protocol that enables Alice and Bob to share a
key and then discuss the effect Eve has on the states received by Bob. We
will assume that Eve uses an intercept-and-resend attack and calculate error
rates that allow the legitimate parties to detect her presence. There are
other more sophisticated forms of attack available to Eve but we will not
analyse them here; we simply remark that this form of attack provides a
useful guide to the security of the protocol against more general attacks.

We first present the \emph{highly-sensitive-to-eavesdropping (HSE)} protocol
in its general form; encoding an alphabet, $\mathcal{A}$, containing $%
c\equiv |\mathcal{A}|$ elements using $d$ dimensional quantum systems. In
Sec. \ref{Example3}, we give an explicit example of the protocol when used
to encode a 4-letter alphabet, say $\{0,1,2,3\},$ using 3-dimensional
quantum systems. A further example is provided in Sec. \ref{alt six state}
where we discus the case of $c=3$ and $d=2$ in an experimental setting.

\paragraph{The HSE-Protocol}

\begin{itemize}
\item Alice and Bob agree publicly on a method of encoding the $c$ elements
of $\mathcal{A}$ using states in $\mathbb{C}^{d}$ by choosing bases $%
\mathcal{B}^{x}=\{|\psi _{i}^{x}\rangle \in \mathbb{C}^{d}:i=1\ldots d\}$
for all $x\in \mathcal{A}.$ They are free to choose any bases provided they
are different in the sense that no two bases have any state in common. We
will discuss the optimum choice in Sec. \ref{distance}. Throughout the
protocol, Alice and Bob will use bases chosen from the set $\{B^{x}:x\in 
\mathcal{A}\}.$

\item Alice generates a random string, $S$, of letters from $\mathcal{A}$
that form the raw data she will attempt to share with Bob.

\item For each element, $x\in S$, Alice generates $c-1$ random numbers, $%
\mathbf{a}\equiv (a_{1},\ldots a_{c-1})$, between $1$ and $d.$ The numbers $%
\mathbf{a}$ serve as indices for states chosen from basis $\mathcal{B}^{x}$
as she now prepares and sends the $c-1$ states $|\psi _{a_{k}}^{x}\rangle
\in \mathcal{B}^{x}$, $k=1\ldots c-1$, to Bob.

\item Bob chooses a sequence of $c-1$ \emph{different} letters of $\mathcal{A%
}$, $x_{1},\ldots ,x_{k}$. When he receives the $k$th state, $|\psi
_{a_{k}}^{x}\rangle ,$ he measures it in the bases $\mathcal{B}^{x_{k}}$ and
records the measurement outcomes, $\mathbf{b}\equiv (b_{1},\ldots b_{c-1}).$

\item After Bob's measurements, Alice publicly announces the indices $%
\mathbf{a}$ keeping her choice of basis a secret. Using this information,
Bob is (sometimes) able to deduce which basis Alice used and therefore to
determine the element of $S$.

\item Bob tells Alice for which elements he was able to determine $x$.
Unsuccessful attempts are discarded, leaving only the shared key. 
\end{itemize}

An element $x\in S$ is successfully shared between Alice and Bob when for
every state $|\psi _{a_{k}}^{x}\rangle \in \mathcal{B}^{x}$, $k=1\ldots c-1,$
the index measured by Bob does \emph{not} equal the index announced by
Alice, $a_{k}\neq b_{k}$ for all $k.$ If this happens Bob knows that \emph{%
none} of his measurements were in basis $\mathcal{B}^{x}$ and so his missing
basis corresponds to the correct letter of the string, $x.$ If Bob's
measurement does equal the announced index for any $k$, he does not know if
this was because he measured in the same basis as Alice or because of the
non-zero overlap between vectors from different bases. This element of the
string then fails.

The protocol presented in \cite{khan+09} is then a special case of this
protocol applied to a two-letter alphabet $\{0,1\}$ so that Alice needs only
to send one state for each letter of $S$. Khan et al.'s protocol is
interesting because it has a high error rate that approaches 50\% for higher
dimensional quantum systems if Alice and Bob use two mutually unbiased
bases. Starting with the probability that the transmission of the element $x$
is successful, we will analyse the performance of the general protocol in
the following sections. We find that this general protocol has an error rate
that approaches 100\% when Alice and Bob use high-dimensional systems and a
complete set of $(d+1)$\ mutually unbiased bases. In Sec. \ref{distance}\ we
will use a natural measure of distance between bases to argue that the
optimal settings for Alice and Bob are indeed mutually unbiased bases.

\subsection{A four-letter alphabet encoded using qutrits\label{Example3}}

We now make the protocol explicit when applied to a four-letter alphabet,
say $\mathcal{A}=\{0,1,2,3\}$, encoded using three-dimensional quantum
systems. Note that we can think of $0,1,2,3$ as representing $00$, $01$, $10$%
, $11$ and therefore the key that Alice and Bob share as pairs of bits, for
example, the string $S=213101$ becomes $100111010001$; this makes it easier
to compare the bit efficiency of different protocols. We examine the case
where Alice and Bob encode $\mathcal{A}$ using the bases 
\begin{eqnarray}
&&\mathcal{B}^{0}\simeq \left( 
\begin{array}{ccc}
1 & 0 & 0 \\ 
0 & 1 & 0 \\ 
0 & 0 & 1%
\end{array}%
\right) ,\mathcal{B}^{1}\simeq \frac{1}{\sqrt{3}}\left( 
\begin{array}{ccc}
1 & 1 & 1 \\ 
1 & \omega  & \omega ^{2} \\ 
1 & \omega ^{2} & \omega 
\end{array}%
\right) ,  \notag \\
&&\mathcal{B}^{2}\simeq \frac{1}{\sqrt{3}}\left( 
\begin{array}{ccc}
1 & 1 & 1 \\ 
\omega ^{2} & 1 & \omega  \\ 
\omega ^{2} & \omega  & 1%
\end{array}%
\right) ,\mathcal{B}^{3}\simeq \frac{1}{\sqrt{3}}\left( 
\begin{array}{ccc}
1 & 1 & 1 \\ 
\omega  & \omega ^{2} & 1 \\ 
\omega  & 1 & \omega ^{2}%
\end{array}%
\right) ,  \label{4 MUBs}
\end{eqnarray}%
where the columns of the matrix $\mathcal{B}^{x}$ correspond to the vectors $%
|\psi _{i}^{x}\rangle ,i=1,2,3$ of each basis. 

In order to send the first element of the string, say $x=2$, Alice generates
three random numbers $a_{1},a_{2},a_{3}\in \{1,2,3\}$ and sends the states $%
|\psi _{a_{1}}^{2}\rangle $, $|\psi _{a_{2}}^{2}\rangle $ and $|\psi
_{a_{3}}^{2}\rangle $. Bob now measures in three different, randomly chosen
bases resulting in the measurement outcomes $b_{1},b_{2}$ and $b_{3}.$ The
element $x$ is successfully transmitted if and only if $a_{1}\neq b_{1},$ $%
a_{2}\neq b_{2}$ and $a_{3}\neq b_{3}$ since if this happens, Bob can be
certain that he did not use the same basis as Alice. Bob must have performed
measurements in the bases $\mathcal{B}^{0},$ $\mathcal{B}^{1}$ and $\mathcal{%
B}^{3}$ so that his missing basis corresponds to the correct element $x=2.$
The probability that an element is shared for each run of the protocol is
given by%
\begin{equation*}
\mathcal{R}_{s}\equiv \frac{1}{4}\left( 1-\frac{1}{3}\right) ^{3}=\frac{2}{27%
},
\end{equation*}%
since there is a $1/4$ chance that Bob does \emph{not} use $\mathcal{B}^{2}$
and a $2/3$ chance that he does \emph{not} measure index $a_{k}$ when using
basis $\mathcal{B}^{x},$ $x\neq 2,$ for $k=1\ldots 3.$

Each element of the string represents two bits and so, on average, in order
to share one bit of information Alice and Bob need to perform this procedure 
$27/4\approx 7$ times so that Alice has to send a total of $3\times
27/4\approx 20.3$ states. This is relatively high, for example in the BB84
protocol, Alice needs to send an average of only two states in order to
successfully transmit one bit of information. However, as we will see in Sec %
\ref{Eve's strategy}, the presence of an eavesdropper causes a much higher
error rate. The present protocol therefore remains secure even if there is a
very high level of system noise.

\subsection{Probability of success}

Having calculated the success rate for the protocol in the case of a
four-letter alphabet encoded using a specific choice of bases of $\mathbb{C}%
^{3}$, we now consider the general probability of success. The protocol
results in a letter, $x,$ forming part of the shared key whenever the
indices measured by Bob are all different from those announced by Alice,
that is whenever $a_{k}\neq b_{k}$ for $k=1,\ldots ,c-1.$ For each state,
indexed by $k,$ Bob makes a measurement in basis $B^{x_{k}}$ so that the
probability of measuring index $a_{k}$ is given by 
\begin{equation*}
q_{k}\equiv \text{prob}(a_{k}=b_{k})=\left\vert \langle \psi
_{a_{k}}^{x_{k}}|\psi _{a_{k}}^{x}\rangle \right\vert ^{2}.
\end{equation*}%
Hence the success rate of the protocol is 
\begin{equation}
\mathcal{R}_{s}\equiv \frac{1}{c}\prod_{k=1}^{c-1}(1-q_{k}),
\label{prod success}
\end{equation}%
the chance that none of the $c-1$ bases chosen by Bob equal the one selected
by Alice, $\mathcal{B}^{x},$ multiplied by the probability of never
measuring the same index even though all of Bob's measurements are different
to $\mathcal{B}^{x}.$ In order to get a success rate \emph{per bit} of
information shared between Alice and Bob, called the \emph{bit} \emph{%
transmission rate,}%
\begin{equation}
\mathcal{R}_{t}\equiv \log _{2}(c)\mathcal{R}_{s},  \label{trans rate}
\end{equation}%
we multiply $\mathcal{R}_{s}$ by $\log _{2}(c)$.

This general formula depends on the choice of bases used to encode the
alphabet, and in particular the modulus of the overlap between states from
different bases. We will consider different bases used in the protocol in
Sec. \ref{specific bases} and compare the bit transmission rate, $\mathcal{R}%
_{t},$ with existing QKD protocols in the conclusion.

\section{Error rate introduced by an eavesdropper \label{Eve's strategy}}

We have seen how the protocol allows Alice and Bob to create a shared key,
we now consider the effect of an eavesdropper. In particular, we analyse the
effect of an intercept-and-resend attack. That is, for each state sent by
Alice, an eavesdropper performs a measurement\ on the system and then
prepares and sends a new state to Bob. In effect, we can imagine the attack
as being performed in two stages. Eve measures the state of the system and
then discards it completely. Using the classical information corresponding
to her measurement outcome, she then prepares a new system in a state that
is as \textquotedblleft close as possible\textquotedblright\ to the original.

In general, Eve is free to use different measurements for each state sent by
Alice. She can also send Bob a system in any state regardless of the
measurement outcome. However, since the states $|\psi _{a_{k}}^{x}\rangle $
have indices, $a_{k},$ that are uniformly distributed, each subsequent
measurement made by Eve is independent from the previous measurement
outcomes. Therefore, there is no loss of generality in assuming that Eve
always uses the same measurement basis, $\mathcal{E}=\{|e_{i}\rangle \in 
\mathbb{C}^{d},i=1\ldots d\},$ corresponding to her optimal one. In
addition, we assume that Eve sends the state corresponding to her
measurement outcome since it is likely to be the state closest to $|\psi
_{a_{k}}^{x}\rangle .$

Alice and Bob can detect Eve's attack in one of two different ways; by
detecting a change in the index of the state received by Bob, called the 
\emph{index transmission error rate} (ITER); and by errors in the final
shared key, called the \emph{quantum bit error rate} (QBER). We begin by
considering the ITER, which can be detected whenever Alice and Bob use the 
\emph{same} bases, $\mathcal{B}^{x},$ and has been used in other QKD
protocols to detect an eavesdropper \cite{khan+09, beige+02a, beige+02b}.

\subsection{The index transmission error rate}

Suppose Alice sends the state $|\psi _{i}^{x}\rangle $, Bob can detect Eve
if he happens to perform a measurement in basis $\mathcal{B}^{x}$ and his
measurement outcome, $j$, does not equal $i$. This occurs with probability $%
p_{i}(x,x)$, where we define 
\begin{equation}
p_{i}(x,y)\equiv \sum_{k=1}^{d}\sum_{\substack{ j=1 \\ j\neq i}}^{d}|\langle
\psi _{i}^{x}|e_{k}\rangle |^{2}|\langle e_{k}|\psi _{j}^{y}\rangle |^{2},
\label{prob}
\end{equation}%
to be the probability that the index $i$ changes when Alice prepares a state
in basis $\mathcal{B}^{x}$ and Bob measures the system he receives in basis $%
\mathcal{B}^{y}$. Since for any $y$ and $k$, Eve measures one of the
possible outcomes with certainty, 
\begin{equation}
\sum_{j=1}^{d}|\langle e_{k}|\psi _{j}^{y}\rangle |^{2}=1,
\label{sum componets}
\end{equation}%
Eqn. (\ref{prob}) can be written as 
\begin{equation*}
p_{i}(x,y)=1-\sum_{k=1}^{d}|\langle \psi _{i}^{x}|e_{k}\rangle |^{2}|\langle
e_{k}|\psi _{i}^{y}\rangle |^{2},
\end{equation*}%
one minus the probability that Bob measures a state with index $i$.

The rate at which Alice and Bob can detect an index transmission error, $%
\mathcal{R}_{IT},$ is calculated by averaging $p_{i}(x,x)$ over all indices, 
$i,$ and letters of the alphabet, $x\in \mathcal{A}$. That is,  
\begin{eqnarray}
\mathcal{R}_{IT} &\equiv &\frac{1}{cd}\sum_{x=0}^{c-1}%
\sum_{i=1}^{d}p_{i}(x,x)  \notag \\
&=&1-\frac{1}{cd}\sum_{x=0}^{c-1}\sum_{i=1}^{d}\sum_{k=1}^{d}|\langle \psi
_{i}^{x}|e_{k}\rangle |^{4}.  \label{ITER}
\end{eqnarray}%
As with the probability of success, $\mathcal{R}_{IT}$ depends on the choice
of bases. We will see how this measure of the sensitivity of the protocol to
eavesdropping can be understood as a measure of distance between the bases
used by all three parties in Sec. \ref{distance}. Then in Sec. \ref{specific
bases} we will consider some interesting examples of specific bases.

\subsection{The quantum bit error rate}

In addition to the index transmission error rate, Alice and Bob can detect
an eavesdropper by calculating the error rate of the final shared key. Eve's
intercept-and-resend attack may cause a change in the index in such a way
that Bob adds an incorrect letter to his key. Just as in the original BB84
protocol, the legitimate parties can detect quantum bit errors by selecting
a random subset of the key and openly comparing its elements.

To see how an error in the key is created, suppose Alice attempts to share
the letter $x\in \mathcal{A}$. If none of the indices measured by Bob equal
the indices announced by Alice, $a_{k}\neq b_{k}$ for all $k=1\ldots c-1$,
Alice adds $x$ to her key and Bob adds $\widetilde{x}$. The letters, $x$ and 
$\widetilde{x}$, correctly coincide provided one of Bob's measurements was
not in the basis $\mathcal{B}^{x}$ since he adds the letter corresponding to
his missing basis. If however, Bob did use $\mathcal{B}^{x}$, he adds the
letter $\widetilde{x}\neq x$ to his key and there is an error in the shared
key. Therefore, the proportion of key elements that contain an error, is
given by the \emph{quantum bit error rate} 
\begin{equation}
\mathcal{R}_{QB}\equiv \frac{c-1}{c}\frac{\mathcal{R}_{BE}}{\mathcal{R}_{K}},
\label{QBER}
\end{equation}%
where; the factor $\frac{c-1}{c}$ is the probability that Bob uses the same
basis as Alice in one of his $c-1$ measurements; $\mathcal{R}_{BE}$ is the
rate at which Bob adds incorrect letters to his key, called \emph{Bob's
error rate}; and $\mathcal{R}_{K}$ is the average probability that a bit is
added to the key regardless of Bob's choice of basis, called the \emph{key
rate}.

We now calculate the terms in Eqn. (\ref{QBER}) starting with $\mathcal{R}%
_{K}$. Given any vector of indices, $\mathbf{a}=(a_{1},\ldots ,a_{c-1})$,
chosen by Alice and bases with indices $\mathbf{y}=(y_{1},\ldots ,y_{c-1})$
chosen by Bob, the probability that $a_{k}\neq b_{k}$ for all $k=1\ldots c-1$
is given by 
\begin{equation}
\prod_{k=1}^{c-1}p_{a_{k}}(x,y_{k}).  \label{prob product}
\end{equation}%
where $p_{i}(x,y)$ has been defined in Eqn. (\ref{prob}). Alice uses vectors
from the set $I\equiv \{(a_{1},\ldots ,a_{c-1}):a_{k}\in \mathbb{Z}_{d}\}$
since she is free to repeat an index. Bob, however is more restricted, he
must use each basis only once and therefore, choose a vector

\begin{equation*}
\mathbf{y\in }Y\equiv \{(y_{1},\ldots ,y_{c-1}):y_{k}\in \mathcal{A}\text{
and }y_{k}\neq y_{l}\text{ for all }k,l\}.
\end{equation*}%
Hence, $\mathcal{R}_{K}$ is the average over all bases $\mathcal{B}^{x}$ and
elements of the sets $I$ and $Y,$ 
\begin{equation}
\mathcal{R}_{K}=\frac{1}{c|Y||I|}\sum_{x=0}^{c-1}\sum_{\mathbf{y}\in Y}\sum_{%
\mathbf{a}\in I}\prod_{k=1}^{c-1}p_{a_{k}}(x,y_{k}),  \label{IC rate}
\end{equation}%
where $|Y|=c!$ and $|I|=d^{c-1}.$ 

The numerator in Eqn. (\ref{QBER}), $\mathcal{R}_{BE},$ is the average
probability that Bob adds an incorrect letter to his key. Such a bit error
occurs when Bob uses the same basis as Alice \emph{and} measures indices
that are all different to those announced by Alice. To help calculate Bob's
error rate, we define the set $Z$ to be 
\begin{equation*}
Z\equiv \{(x,z_{2},\ldots ,z_{c-1}):z_{k}\in \mathcal{A}\text{, }z_{k}\neq x%
\text{ and }z_{k}\neq z_{l}\text{ for all }k,l\},
\end{equation*}%
that is, the first component of every $\mathbf{z}\in Z$ corresponds to the
letter $x$ used by Alice to encode the states. Therefore, Bob's error rate
is given by%
\begin{equation}
\mathcal{R}_{BE}=\frac{1}{c|Z||I|}\sum_{x=0}^{c-1}\sum_{\mathbf{z}\in
Z}\sum_{\mathbf{a}\in I}\prod_{k=1}^{c-1}p_{a_{k}}(x,z_{k}),  \label{BE rate}
\end{equation}%
where we average over all outcomes that correspond to Bob adding an
incorrect letter to his key and the set $Z$ contains $|Z|=(c-1)!$ elements.

The rather complicated formula for $\mathcal{R}_{QB}$ given by Eqns. (\ref%
{QBER}), (\ref{IC rate}) and (\ref{BE rate}) has a simple form when Alice
and Bob use only two bases in the protocol. The simplification is due to the
fact that when $c=2$, Bob's error rate $\mathcal{R}_{BE}=\mathcal{R}_{IT}$
and hence 
\begin{equation*}
\mathcal{R}_{QB}=\frac{\mathcal{R}_{IT}}{2\mathcal{R}_{K}}\text{ \ \ for }%
c=2,
\end{equation*}%
corresponding to the QBER obtained in \cite{khan+09}$.$ We will also see
that the general form of $\mathcal{R}_{QB}$ simplifies when applied to a
specific choice of bases in Sec. \ref{specific bases}. Before doing so, we
show how the error rate $\mathcal{R}_{IT}$ relates to a natural measure of
distance between the bases of $\mathbb{C}^{d}$ used by the three parties.

\section{ Distance between bases \label{distance}}

In this section we consider the bases used in the QKD protocol as points in
a higher-dimensional space. This setting allows us to understand the optimal
strategy for the legitimate parties in terms of a natural measure of
distance between two bases. We follow an approach similar to that presented
in \cite{bengtsson+07}; here, however, we will consider an alternative
choice of origin so that the resulting space is an \emph{affine} space
rather than a vector space.

We begin by associating to every normalised vector, $|\psi \rangle \in 
\mathbb{C}^{d}$, the operator%
\begin{equation*}
|\psi \rangle \rightarrow \mathbf{\psi }=|\psi \rangle \langle \psi |
\end{equation*}%
that lives in a $d^{2}-1$ dimensional space consisting of Hermitian
operators of trace one. Equipped with the inner product 
\begin{equation*}
\mathbf{\psi \cdot \phi =}\text{Tr}\mathbf{\psi \phi ,}
\end{equation*}%
this is an affine space in which a basis $\mathcal{B}=\{|\psi _{1}\rangle
,|\psi _{2}\rangle ,\ldots ,|\psi _{d}\rangle \mathbf{\}}$ of $\mathbb{C}%
^{d} $ is identified with a set of operators $\{\mathbf{\psi }_{1},\mathbf{%
\psi }_{2},\ldots ,\mathbf{\psi }_{d}\mathbf{\}}$ spanning a $d-1$
dimensional plane. To define a distance between two such planes, we perform
a similar procedure and embed them in an even larger space so that to each
basis $\mathcal{B}$ we associate the matrix

\begin{equation*}
\Psi =\frac{1}{\sqrt{d}}[\mathbf{\psi }_{1}\mathbf{\psi }_{2}\ldots \mathbf{%
\psi }_{d}]\left[ 
\begin{array}{c}
\mathbf{\psi }_{1}^{T} \\ 
\mathbf{\psi }_{2}^{T} \\ 
\vdots  \\ 
\mathbf{\psi }_{d}^{T}%
\end{array}%
\right] ,
\end{equation*}%
that projects onto the plane spanned by the basis vectors $\{\mathbf{\psi }%
_{1},\mathbf{\psi }_{2},\ldots \mathbf{\psi }_{d}\mathbf{\}.}$ Acting on an
arbitrary pure state, $\mathbf{\phi ,}$ the operator $\Psi $ describes the
action of performing a measurement in basis $\mathcal{B}$ since the non-zero
elements of $\Psi \mathbf{\phi }$ are $\left\vert \langle \psi _{i}|\phi
\rangle \right\vert ^{2}\mathbf{\psi }_{i}$ for $i=1\ldots d.$

The matrices $\Psi $ are elements of a $(d^{2}-1)^{2}$ dimensional space
(called an affine Grassmannian), in which a natural measure of distance
between two points, $\Phi $ and $\Psi ,$ is the chordal Grassmannian distance%
\begin{equation}
D^{2}(\Phi ,\Psi )=1-\text{Tr}\Phi \Psi .  \label{distance measure}
\end{equation}%
Applying this distance measure to two points, $\Phi $ and $\Psi ,$
associated with bases reads 
\begin{eqnarray*}
D^{2}(\Phi ,\Psi ) &=&1-\frac{1}{d}\text{Tr}\left\{ [\mathbf{\psi }_{1}%
\mathbf{\psi }_{2}\ldots \mathbf{\psi }_{d}]\left[ 
\begin{array}{c}
\mathbf{\psi }_{1}^{T} \\ 
\mathbf{\psi }_{2}^{T} \\ 
\vdots \\ 
\mathbf{\psi }_{d}^{T}%
\end{array}%
\right] [\mathbf{\varphi }_{1}\mathbf{\varphi }_{2}\ldots \mathbf{\varphi }%
_{d}]\left[ 
\begin{array}{c}
\mathbf{\varphi }_{1}^{T} \\ 
\mathbf{\varphi }_{2}^{T} \\ 
\vdots \\ 
\mathbf{\varphi }_{d}^{T}%
\end{array}%
\right] \right\} \\
&=&1-\frac{1}{d}\sum_{i=1}^{d}\sum_{j=1}^{d}(\mathbf{\psi }_{i}\mathbf{\cdot
\varphi }_{j})^{2} \\
&=&1-\frac{1}{d}\sum_{i=1}^{d}\sum_{j=1}^{d}\left\vert \langle \psi
_{i}|\varphi _{j}\rangle \right\vert ^{4}.
\end{eqnarray*}%
Hence the average distance, $D_{average}$, between Eve's basis $E$ and the
bases chosen by Alice and Bob, $B^{x},$ $x=0\ldots c-1,$ is given by%
\begin{eqnarray}
D_{average} &=&\frac{1}{c}\sum_{x=0}^{c-1}D^{2}(B^{x},E)  \notag \\
&=&1-\frac{1}{cd}\sum_{x=0}^{c-1}\sum_{i=1}^{d}\sum_{k=1}^{d}|\langle
e_{k}|\psi _{j}^{x}\rangle |^{4}  \label{Av distance} \\
&=&\mathcal{R}_{IT},  \notag
\end{eqnarray}%
the index transmission error rate caused by Eve's intercept-and-resend
attack.

This distance measure provides an intuitive feel as to how the three parties
in the protocol should behave: Alice and Bob aim to maximize the error rate $%
\mathcal{R}_{IT}$ by separating their bases as much as possible; whilst Eve
chooses a basis that minimises the average distance between all of the bases
chosen by Alice and Bob. We will begin the next section by making these
statements more precise and find that they lead to the conclusion that Alice
and Bob should use a complete set of mutually unbiased bases.

\section{Optimal choice of bases \label{specific bases}}

In this section we consider specific choices of bases used by Alice and Bob
in the HSE-protocol. The protocol is entirely general and any set of bases
can be used to encode the alphabet. There are likely to be many
considerations in choosing a suitable set such as the ease of preparing and
measuring states in\ each of the prescribed bases. In this section we will
not worry about experimental difficulties but simply consider the optimal
choice from a theoretical perspective. Motivated by the distance measure in
Sec. \ref{distance} we begin by considering a set of mutually unbiased (MU)
bases.

\subsection{Mutually unbiased bases}

Two bases $\mathcal{B}^{x}=\{|\psi _{i}^{x}\rangle ,i=1\ldots d\}$ and $%
\mathcal{B}^{y}=\{|\psi _{i}^{y}\rangle ,i=1\ldots d\}$ are called mutually
unbiased if the modulus of the inner product of vectors from different bases
is uniform, 
\begin{equation}
|\langle \psi _{i}^{x}|\psi _{j}^{y}\rangle |=\kappa ,  \label{MUBs}
\end{equation}
which in finite dimensions means that $\kappa =1/\sqrt{d}$. Schwinger noted 
\cite{schwinger60} that two such bases represent measurements that are
\textquotedblleft maximally non-commuting\textquotedblright\ in that
measuring in one bases reveals no information about the outcome of a
measurement in the other basis. For example, in dimension $d=2$, if we set a
Stern-Gerlach experiment to measure spin in the $x$ direction, we gain no
information about spin in the $z$ direction.

This maximal lack of information about measurement outcomes from other bases
is captured by the distance measure introduced in Eqn. (\ref{distance
measure}). The distance between any two bases $\Phi $ and $\Psi ,$ is
bounded by 
\begin{equation*}
0\leq D^{2}(\Phi ,\Psi )\leq 1-\frac{1}{d},
\end{equation*}%
where the lower bound is obtained when $\Phi $ and $\Psi $ span the same
subspace and the upper bound is realised when they are mutually unbiased.
Since Alice and Bob wish to maximize the average distance between all of the
bases they use, a natural strategy is to use as many MU bases as possible.
They cannot use more than $d+1$, called a \emph{complete set}, since it is
impossible to fit any more $d-1$ dimensional planes with the correct overlap
into a space of dimension $d^{2}+1$ \cite{bengtsson+07}. In dimension $d=3$,
the four bases given in Eqn (\ref{4 MUBs}) constitute a complete set of MU
bases. In all other prime-power dimensions, a complete set of MU bases has
been constructed \cite{wootters+89}. However, for composite dimensions $%
d=6,10,12,...$ the maximum number of bases satisfying the conditions (\ref%
{MUBs}), remains an open problem.

We now turn our attention to the optimal strategy of an eavesdropper. As
before, we assume that she uses an intercept-and-resend attack and following
the arguments of Sec. \ref{Eve's strategy}, only uses one basis
corresponding to her optimal choice. Eve's optimal strategy is essentially a
minimisation problem subject to some constraints. The functions she wishes
to minimise are the error rates $\mathcal{R}_{QB}$ and $\mathcal{R}_{IT},$
and the constraints come from the fact that Eve must use a set of $d$
orthonormal vectors. By approaching this problem numerically, Khan et. al.
provide evidence that for $c=2$, the index transmission error rate has a
global minimum when Eve's basis spans the same subspace as one of the bases
chosen by Alice and Bob \cite{khan+09}. In other words, Eve's optimal
strategy is to simply pick one of the bases used by the legitimate parties.\ 

Eve has many alternative eavesdropping strategies at her disposal. For
example, for the case when $d=c=2$, Eve could use the so-called Breidbart
basis that is halfway between the two bases used by the legitimate parties 
\cite{bennett+82}. In the BB84 protocol, such a strategy has been shown to
increase the chance that Eve reads the bit correctly although it does not
reduce her chance of being detected \cite{bennett+92}. However, when the
legitimate parties use a complete set of MU bases, there is no basis that is
\textquotedblleft halfway\textquotedblright\ between all of them. There are
many issues concerned with finding the optimal strategy of an eavesdropper 
\cite{ekert+94,huttner+94,fuchs+97,bechmann+99}. In the following, we will
assume that Eve picks one of the bases used by the legitimate parties and
consider the protocol when Alice and Bob use a set of $c$ MU bases.

There is no loss of generality in assuming that Eve's basis is given by $%
\mathcal{E}\equiv \{|e_{i}\rangle \in \mathbb{C}^{d},i=1\ldots d\}=\mathcal{B%
}^{0}$. Under this assumption, the distance between the bases used by all
three parties is 
\begin{equation*}
D^{2}(E,B^{x})=\left\{ 
\begin{array}{ll}
0\text{ \ } & \text{if }x=0 \\ 
1-\frac{1}{d}\text{ \ \ \ } & \text{if }x\neq 0,%
\end{array}%
\right.
\end{equation*}%
zero if $\mathcal{B}^{x}$ corresponds to $\mathcal{E}$ or maximal otherwise.
Hence, the index transmission error rate for a set of $c$ MU bases is given
by 
\begin{eqnarray}
\mathcal{R}_{IT}^{MUB} &=&\frac{1}{c}\sum_{x=1}^{c-1}\left( 1-\frac{1}{d}%
\right) \text{ \ }  \notag \\
&=&\frac{(c-1)(d-1)}{cd}.  \label{MUBs ITER}
\end{eqnarray}%
%
%
%
%
%
%
%
%
%
%
%
%
%
We see that the error rate is an increasing function of both $c$ and $d$ and
that Eqn. (\ref{MUBs ITER}) is indeed maximized if Alice and Bob use a
complete set of MU bases. In which case, the index transmission error rate
of the protocol equals 
\begin{equation*}
\mathcal{R}_{IT}^{MUB}=\frac{d-1}{d+1}
\end{equation*}%
and therefore tends to 100\% as $d$ tends to infinity.

The index transmission error rate introduced by an intercept-and-resend
attack in Eqn. (\ref{MUBs ITER}) is equal to the quantum bit error rate of
the BKB01-protocol of Bourennane et al. \cite{bourennane+01}. It is a
natural generalisation of the BB84 protocol and has been further analysed in 
\cite{bourennane+02, cerf+02}. The BKB01-protocol, the $d$ letters of an
alphabet are encoded into the indices of one of $c$ mutually unbiased bases.
Alice sends a state $|\psi _{x}^{a}\rangle ,$ where $x=1\ldots d$ and $%
a=0\ldots c-1,$ and after Bob's measurement, announces the basis, $a$, which
she used to prepare the states. Hence, whenever Bob performs a measurement
in the same basis $\mathcal{B}^{b},$ they share the letter $x\in \mathcal{A}$%
. Note that in contrast to the HSE-protocol, the roles of $c$ and $d$ are
reversed. In the conclusion, the error rates and the number of states needed
to successfully share one bit of the key for the BKB01 protocol are compared
to the HSE-protocol.

The quantum bit error rate, $\mathcal{R}_{QB}$, given in Eqn. (\ref{QBER}),
also simplifies significantly when Alice and Bob use a set of $c$ MU bases
and we assume that Eve's basis equals $\mathcal{E=B}^{0},$ say. Under these
assumptions, the probability that an index changes is zero if all three
parties use the same bases and one minus the probability of measuring the
correct index if any one of the parties uses a different basis 
\begin{equation*}
p_{i}(x,y)=\left\{ 
\begin{array}{ll}
0\text{ \ } & \text{if }(x,y)=(0,0) \\ 
1-\frac{1}{d}\text{ \ \ } & \text{if }(x,y)\neq (0,0).%
\end{array}%
\right. 
\end{equation*}%
Therefore, the product of probabilities given in Eqn. (\ref{prob product}), 
\begin{equation*}
\prod_{k=1}^{c-1}p_{a_{k}}(x,y_{k}),
\end{equation*}%
depends solely on whether any of the terms correspond to $(x,y_{k})=(0,0).$

To calculate the number of non-zero terms in $\mathcal{R}_{BE},$ given in
Eqn. (\ref{BE rate}), note that one of Bob's bases is always equal to $%
\mathcal{B}^{x}$ and therefore $(x,y_{k})=(0,0)$ for some $k$, if and only
if $x=0.$ Hence, the proportion of non-zero terms in Eqn. (\ref{BE rate}) is
equal to $\left( 1-1/c\right) $. Similarly, when the vectors $\mathbf{y\in }Y
$, the number of non-zero terms in Eqn. (\ref{IC rate}) is $\left( 1-\frac{1%
}{c}+\frac{1}{c^{2}}\right) |Y||I|c$, so that Bob's error rate and the key
rate are given by 
\begin{eqnarray*}
\mathcal{R}_{BE} &=&\left( 1-\frac{1}{c}\right) \left( 1-\frac{1}{d}\right)
^{c-1} \\
\mathcal{R}_{K} &=&\left( 1-\frac{1}{c}+\frac{1}{c^{2}}\right) \left( 1-%
\frac{1}{d}\right) ^{c-1},
\end{eqnarray*}%
respectively. Therefore, for a set of $c$ mutually unbiased bases, the error
rate $\mathcal{R}_{QB}$ is given by 
\begin{equation}
\mathcal{R}_{QB}^{MUB}=\left( 1-\frac{1}{c}\right) ^{2}\left( 1-\frac{1}{c}+%
\frac{1}{c^{2}}\right) ^{-1}  \label{MUBs QBER}
\end{equation}%
which, surprisingly, does \emph{not} depend on the dimension of the quantum
systems used in the protocol. However, it is of course limited by the number
of MU bases that can be constructed in a given dimension $c\leq d+1$ and may
also be limited by the conjectured non-existence of complete sets of MU
bases in composite dimensions.

Whilst constructions of complete sets of MU bases are known for prime power
dimensions, and are well understood in low dimensions \cite{brierley+09b}
their existence is an open problem for composite dimensions$.$ In fact,
there is considerable numerical \cite{butterley+07, brierley+08} and
analytical \cite{brierely+09, jamming+09} evidence to suggest that there are
no more than three MU bases in dimension six. Hence restricting the
measurements to MU bases could mean that the protocol is more efficient in
prime power dimensions than in composite dimensions, for example, using six
MU bases in dimension five the error rate $\mathcal{R}_{IT}$ is $2/3\approx
66.7\%$ were as if only three MU bases are available in dimension six the
maximum error rate is $5/9\approx 55.6\%.$ The situation for the QBER is
even more pronounced since $\mathcal{R}_{QB}^{MUB}$ depends only on the
number of MU bases available and not on the dimension. As such it would be
better to use quantum systems of dimension three since it is possible to
construct four MU bases than to use systems of dimension $d=6,$ for which we
only know how to construct three bases with the required overlap.

\subsection{Approximate mutually unbiased bases \label{AMUBs}}

It is not clear that a complete set of $d+1$ mutually unbiased bases exists
in all dimensions. Therefore, in order to consider the limiting behaviour of
the protocol, we consider an alternative choice of bases for which
constructions are known in all dimensions. As with a complete set of MU
bases, they have the property that the error rate $\mathcal{R}_{IT}$ tends
to 100\% as the dimension of the quantum systems used by Alice and Bob
increases.

By relaxing the uniform modulus condition (\ref{MUBs}), Klappenecker et al. 
\cite{klappenecker+05}\ define approximate mutually unbiased bases
(abbreviated as AMU bases) which have the property that the modulus of the
inner product between vectors from different bases is small. In particular
they define a set of $d^{2}$ bases such that 
\begin{equation*}
|\langle \psi _{i}^{x}|\psi _{j}^{y}\rangle |\leq \frac{2+O(d^{-1/10})}{%
\sqrt{d}}\text{ \ for }x\neq y,
\end{equation*}%
and for all $i,j$, where $f(d)=O(d^{-1/10})$ means that there exists a
constant $K>0$ such that $|f(d)|\leq Kd^{-1/10}$ for all $d\geq 1.$ Hence if
Alice and Bob use all $d^{2}$ bases, and Eve uses one of the bases in her
intercept-and-resend attack, the index transmission error rate is bounded
from below by 
\begin{equation}
\mathcal{R}_{IT}^{AMUB}\geq 1-\frac{1}{d^{3}}\left[
d+(d^{2}-1)(2+Kd^{-1/10})^{4}\right] .  \label{AMUBS error rate}
\end{equation}%
The unknown constant in Eqn. (\ref{AMUBS error rate}) prevents us from
saying anything in specific dimensions, but we can still consider the
protocol when Alice and Bob use a set of AMU bases in the limit as $d$ tends
to infinity. We see that such a set of approximate MU bases defined so that
they minimise the value of $\kappa $ in Eqn. (\ref{MUBs}) and therefore
maximise the distance measure defined by Eqn. (\ref{Av distance}) are good
at detecting the eavesdropping by Eve. Even though a complete set of MU
bases may not exist in every dimension, we can at least define a set of AMU
bases that do exist in all dimensions and for which the ITER tends to $100\%$%
.

\section{Implementations\label{Example}}

In this section, we present a specific example of how Alice and Bob can use
the HSE-protocol to form a shared key. We also calculate the quantum bit and
index transmission error rates that allow Alice and Bob to detect an
eavesdropper for this choice of $c$ and $d$. Finally, we discuss a practical
implementation of the protocol that could be used for any values of $c$ and $%
d$ using photon states and multiport beam splitters.

\subsection{An alternative \textquotedblleft six-state\textquotedblright\
protocol using qubits \label{alt six state}}

In the six-state protocol \cite{bechmann+99,bruss98}, Alice prepares and
sends one of six states corresponding to the points on the Bloch ball $(\pm
1,0,0),$ $(0,\pm 1,0)$ and $(0,0,\pm 1)$. These six states form three MU
bases $\mathcal{B}^{0}$, $\mathcal{B}^{1}$, and $\mathcal{B}^{2}$
corresponding to 
\begin{equation*}
\{|0\rangle ,|1\rangle \},\{\frac{1}{\sqrt{2}}(|0\rangle +|1\rangle ),\frac{1%
}{\sqrt{2}}(|0\rangle -|1\rangle )\},\text{ and }\{\frac{1}{\sqrt{2}}%
(|0\rangle +i|1\rangle ),\frac{1}{\sqrt{2}}(|0\rangle -i|1\rangle )\}
\end{equation*}%
respectively. After receiving a state from Alice, Bob performs a measurement
in one of the three bases and records his outcome. Alice announces which of
the bases she used to prepare the state and if Bob used the \emph{same}
basis they are able to share an element of the key. When the bases used by
Alice and Bob coincide, Bob can correctly determine the letter because his
measurement outcome must correspond to the state prepared by Alice (in the
absence of an eavesdropper).

Using the polarization of photons to encode the states, Enzer et. al. have
implemented the six-state protocol experimentally \cite{enzer+02}. The three
bases in their scheme correspond to horizontal/vertical (H/V), diagonal $%
+45^{\circ }$/$-45^{\circ }$ (D/d) and left/right circular (L/R)
polarization; the three states H,D and L encoding a zero and V,d,R a one. By
simulating an intercept-and-resend attack Enzer et. al. find a bit error
rate of $34.0\pm 1.4\%$ in agreement with the theoretical value of $33.3\%$.

In order to implement the HSE-protocol, the six-state scheme presented in 
\cite{enzer+02} requires only a slight modification. The preparation and
measurement of the states remains the same; the difference being the method
of encoding the alphabet. Here, we will use the polarizations H/V to encode
a zero, D/d a one and L/R a two. That is, our scheme uses a three letter
alphabet $\mathcal{A}=\{0,1,2\}$ encoded into the choice of basis; $\mathcal{%
B}^{0}$, $\mathcal{B}^{1}$, or $\mathcal{B}^{2}.$ The indices of the states
are either zero or one corresponding to H,D, and L or V,d and R respectively.

As before, Alice chooses one of the bases $\mathcal{B}^{0}$, $\mathcal{B}%
^{1} $, or $\mathcal{B}^{2}$ but this time sends \emph{two} states. That is,
suppose Alice chooses to encode the bits in the H/V basis, then she sends
either HH, HV, VH or VV. Bob now makes a measurement in two \emph{different}
bases and records the indices corresponding to his outcomes. Alice announces
the indices, either 00, 01, 10 or 11, equal to her choice of prepared
states. She does not announce the basis. Using the indices announced by
Alice and his measurement outcomes, Bob hopes to determine the basis used by
Alice.

An element of the key is shared whenever Bob's indices both differ from the
indices announced by Alice. For example, if Alice sends states with indices
01, an element of the key is shared if and only if Bob's measurement
outcomes are 10. For this scheme, the average rate at which a bits are
shared between Alice and Bob is given by 
\begin{equation*}
\mathcal{R}_{t}=\log _{2}(3)\frac{1}{3}\left( 1-\frac{1}{2}\right)
^{2}\approx 13.2\%,
\end{equation*}%
since the probability that Bob does not use the same basis as Alice in both
of his measurements is $1/3$ and there is a $1/2$ chance that he does not
measure the announced index when using a different basis. The pre-factor of $%
\log _{2}(3)$ is due to the fact that when an element of the key is shared
it corresponds to an element of of a \emph{three} letter alphabet.

Whilst the bit rate is $13.2\%$, compared to $33\%$ for the six-state
protocol, the number of sates Alice must send in order to share one bit of
information is much higher than in the six-state protocol. For each attempt
at sharing a letter of the alphabet, Alice must send two states. Therefore
the average number of states, $\mathcal{N}_{s}=2\times 100/13.2\approx 15.2$
which is five times more than the $3$ states needed to share one bit when
implementing the six-state protocol.

We find that although this protocol is more expensive than the six-state
protocol, it is also more sensitive to an eavesdropper. The quantum bit
error rate of an intercept-and-resend attack of this new protocol is given
by 
\begin{equation*}
\mathcal{R}_{QB}^{MUB}=\left( 1-\frac{1}{3}\right) ^{2}\left( 1-\frac{1}{3}+%
\frac{1}{3^{2}}\right) ^{-1}=\frac{4}{7}\approx 57.1\%,
\end{equation*}%
following Eqn. (\ref{MUBs QBER}); representing a significant improvement
over the $33.3\%$ error rate of the six-state protocol. We have used the 
\emph{same} six states as the six-state protocol but this new method of
encoding the letters of an alphabet is more sensitive to an
intercept-and-resend attack.

\subsection{Possible implementation using multiport beam splitters}

A recent experiment has implemented quantum state tomography using a
complete set of MU bases in dimension $d=4$ \cite{adamson+08}. It
demonstrates that tomography with MU bases is not only optimal in theory,
but is more efficient than standard measurement strategies in practice. The
scheme presented in \cite{adamson+08} therefore provides a way of measuring
two-qubit photon states in one of five mutually unbiased bases in dimension
four. However, to implement the QKD presented in Sec. \ref{general scheme} a
set of $c$ MU bases, we also need to reliably \emph{prepare} the relevant
states. Such a scheme for $c=2$ MU bases has been provided by Khan et al. 
\cite{khan+09} and can be extended to any number of mutually unbiased bases.
This follows from the fact that \emph{any} discrete unitary operator can be
realised using a series of beam splitters and mirrors \cite{reck+94}. These
so called \emph{multiport beam splitters} are symmetric when they correspond
to MU bases \cite{zukowski+97}.

The protocol could be implemented as follows. Alice uses a single photon
source such as a spontaneous parametric down conversion crystal. She now
chooses one of $c-1$ multiport beam splitters, or to bypass the beam
splitters altogether. This gives one of the $c$ bases labeled by the letters
of $\mathcal{A}$ required for the protocol. Each vector $|\psi
_{i}^{x}\rangle $ of her chosen basis, $\mathcal{B}^{x}$, is encoded into
the output paths of the corresponding beam splitter by sending a single
photon into the input port $i$. Bob uses the same beam splitters in order to
measure the state of each photon he receives. He does this by first sending
it through one of the beam splitters (or bypasses them to measure $\mathcal{B%
}^{0}$) and then detecting it in one of the $d$ output ports. When $c=2,$ a
natural choice for the two MU bases is to use the standard basis $\mathcal{B}%
^{0}=\{|i\rangle ,i=0\ldots d-1\}$ and the so called Fourier matrix which
has entries $F_{ij}=\omega ^{ij}/\sqrt{d},$ for $i,j=0\ldots d-1$ where $%
\omega =\exp (2\pi i/d)$ is the $d$th root of unity (which for $d=3$ is
given by $\mathcal{B}^{1}$ in Eqn. (\ref{4 MUBs})). This scheme corresponds
to the one presented in \cite{khan+09} and could be realised using Bell
multiport beam splitters \cite{lim+05}.

\section{Conclusion \label{conclusion}}

We have presented a novel protocol that enables two parties to generate a
shared key. It is special in that the presence of an eavesdropper who uses
an intercept-and-resend attack creates a \emph{high error rate}. This has
the practical advantage of allowing Alice and Bob to detect Eve even if the
system noise in their implementation is high. We have analysed two error
rates that allow for the detection of an eavesdropper;\ the \emph{index
transmission error rate} (ITER) and the \emph{quantum bit error rate}
(QBER). Both of these measures of the sensitivity to eavesdropping tend to
one as the parties use more bases to encode the elements of the key and, in
the case of the ITER, if they use higher dimensional systems.

\begin{table}[th]
\begin{center}
\begin{tabular}{ccccccc}
\hline\hline
Protocol & $(d,c)$ &  & $\mathcal{R}_{QB}$ & $\mathcal{R}_{IT}$ & $\mathcal{R%
}_{t}$ & $N_{s}$ \\ \hline
BB84 & $(2,2)$ &  & $25.0\%$ & n/a & $50.0\%$ & 2.0 \\ 
KMB09 & $(2,2)$ &  & $33.3\%$ & $25.5\%$ & $25.0\%$ & 4.0 \\ 
BKB01 (6-state) & $(2,3)$ &  & $33.3\%$ & n/a & $33.3\%$ & 3.0 \\ 
HSE & $(2,3)$ &  & $57.1\%$ & $33.3\%$ & $13.2\%$ & 15.1 \\ 
&  &  &  &  &  &  \\ 
BKB01 & $(3,2)$ &  & $33.3\%$ & n/a & $79.2\%$ & 1.3 \\ 
KMB09 & $(3,2)$ &  & $33.3\%$ & $33.3\%$ & $33.3\%$ & 3.0 \\ 
BKB01 & $(3,4)$ &  & $50.0\%$ & n/a & $39.6\%$ & 2.5 \\ 
HSE & $(3,4)$ &  & $69.2\%$ & $50.0\%$ & $14.8\%$ & 20.3 \\ 
&  &  &  &  &  &  \\ 
BKB01 & $(7,2)$ &  & $42.9\%$ & n/a & $140.4\%$ & 0.7 \\ 
KMB09 & $(7,2)$ &  & $33.3\%$ & $42.9\%$ & $42.9\%$ & 2.3 \\ 
BKB01 & $(7,8)$ &  & $75.0\%$ & n/a & $35.1\%$ & 2.8 \\ 
HSE & $(7,8)$ &  & $86.0\%$ & $75.0\%$ & $12.7\%$ & 54.9 \\ \hline\hline
\end{tabular}%
\end{center}
\caption[Comparison of different QKD protocols in dimensions $d=2,3$ and 7]{%
Table comparing different QKD protocols in dimensions $d=2,3$ and 7; $%
\mathcal{R}_{QB}$ and $\mathcal{R}_{IT}$ are the quantum bit and index
transmission error rates of an intercept-and-resend attack, respectively; $%
\mathcal{R}_{t}$ is the bit transmission rate defined in Eqn (\protect\ref%
{trans rate}); finally, $N_{s}$ is the average number of states Alice must
send in order to share one bit with Bob. Note that the KMB09-protocol is a
special case of the HSE-protocol. }
\label{table}
\end{table}

Table \ref{table} compares the essential features of the HSE-protocol to
existing QKD protocols: the original quantum key distribution protocol of
Bennett and Brassard \cite{bennet+84} is referred to as BB84; the
generalisation of BB84 to a protocol that uses $c$ mutually unbiased bases
and $d$-dimensional quantum systems \cite{bourennane+01} is called BKB01;
the case where three MU bases are used in dimension two corresponds to the
six-state protocol (6-state) \cite{bechmann+99,bruss98}; the protocol
presented in Sec. \ref{general scheme} is denoted HSE (which stands for 
\emph{highly sensitive to eavesdropping}); the case where only two bases are
used corresponding to the protocol of Khan et al. (KMB09) \cite{khan+09}.
Throughout the table, we assume that the HSE-protocol is applied to a set of 
$c$ mutually unbiased bases. The pair of numbers, $(d,c)$, in the second
column correspond to the dimension of the quantum systems used in the
protocol and the number of elements in the classical alphabet.

The third and forth columns of Table \ref{table} show the QBER and the ITER
respectively. The error rates, which have been calculated using Eqns. (\ref%
{MUBs ITER}) and (\ref{MUBs QBER}), show that by using $d+1$ MU bases, Alice
and Bob can increase the QBER beyond that of BKB01. The fifth column
displays the rate at which the two legitimate parties sharing one bit of
information; that is $\mathcal{R}_{s}$ has been normalised so that it gives
a \emph{per bit} success rate\footnote{%
Note that when the BKB01 protocol is applied to two MU bases in dimension $%
d=7$, the rate at which bits are shared between Alice and Bob is larger than
100\%. In this case, the legitimate parties use 7-dimensional quantum
systems so that each time they are successful, they share an element of a 7
letter alphabet. Hence, the number of states Alice needs to send in order to
share one \emph{bit} is 0.7, i.e. less than one.}. The last column then
shows the average number of states Alice needs to send in order to
successfully share one bit of her key with Bob. This final column clearly
demonstrates the trade-off between the error rate and the \textquotedblleft
cost\textquotedblright\ of producing a shared key. It is possible to make it
easier to detect Eve but this comes at the expense of reducing the bit
transmission rate.

At first sight, the protocol appears to have no special features relating to
the dimension of the quantum systems used by Alice and Bob. However, an
analysis of the optimal bases reveals that it is more efficient when the
legitimate parties use systems of prime-power dimensions. In prime-power
dimensions Alice and Bob can use constructions of $d+1$ mutually unbiased
bases that are conjectured not to exist in composite dimensions such as $%
d=6,10,12,$ etc. In addition, in some dimensions, \emph{inequivalent} sets
of $c$ MU bases are available. For example in dimension $d=4$, there exits a
three-parameter family of triples of MU bases \cite{brierley+09b,zauner99}\
or in dimension $d=16$ there is a $17$-parameter family of pairs of MU bases 
\cite{tadej+06}. It may be that within these families there are some bases
that are experimentally more accessible than others. For example, Romero et
al. \cite{romero+05} have considered a notion of inequivalent sets of MU
bases involving the entanglement content of the bases and therefore, one
aspect of the experimental difficulty in measuring and preparing systems in
the corresponding bases.

If an experimenter finds that a particular measurement is easy to implement
and that quantum systems prepared in the corresponding basis are readily
available, they can use the HSE-protocol to distribute shared keys. Given
the analytical form of the bases, we have shown how to calculate the error
rate and the rate at which elements of a key are generated. Hence, to some
extent, the protocol can be made to fit around experimental conditions, the
question is then if the system noise enables an eavesdropper to disguise
their presence. It may be that in practice it is better to search for
measurements that can be performed efficiently in the laboratory (or in a
purpose built device) than to find the analytical optimal bases.

In recent years, quantum physicists have realised that finite dimensional
complex linear spaces are surprisingly rich both in physical content and
from a mathematical perspective. This setting has led to many important
physical discoveries\ and in particular, the ability to distribute keys in a
secure way. In this paper, we have explored this mathematical structure
further and found that, at least in principle, Alice and Bob can make it
very hard for Eve to hide.

\section*{Acknowledgments}

The author would like to thank Stefan Weigert, Tony Sudbery and Almut Beige
for helpful comments on an earlier version of the manuscript.

\end{document}